\documentstyle[twocolumn,prb,aps,epsfig]{revtex}

\begin{document}
\title{
        Fermi surface renormalization in Hubbard ladders }

\author{Kim Louis, J.V. Alvarez, and Claudius Gros 
       } 

\address{Fakult\"at 7, Theoretische Physik,
 University of the Saarland,
66041 Saarbr\"ucken, Germany.}

\maketitle

\begin{abstract}
We derive the one-loop renormalization equations
for the shift in the Fermi-wavevectors for 
one-dimensional interacting models with
four Fermi-points (two left and two right
movers) and two Fermi velocities $v_1$ and
$v_2$. We find the shift to be proportional
to $(v_1-v_2)U^2$, where $U$ is the Hubbard-$U$.
Our results apply to
the Hubbard ladder and to the $t_1-t_2$
Hubbard model. The Fermi-sea with fewer
particles tends to empty. The stability
of a saddle point due to shifts of the
Fermi-energy and the shift of the Fermi-wavevector
at the Mott-Hubbard transition are discussed.
\end{abstract}


\vspace*{1cm}
{\bf Introduction} -
Fermi-surface properties of one- and two-
dimensional interacting electron systems are
a fascinating topic of current interest,
motivated in part by the opening of a 
pseudogap in underdoped cuprates.
Several novel effects have been found
in generalized renormalization-group (RG)
approaches to weak coupling models, in particular the
formation of quasiparticle gaps \cite{fur98}
and a Pomaranchuk \cite{hal00} and charge \cite{gon00}
instability in $N$-patch 2D-Hubbard models \cite{alv98,gon99}.

Novel Fermi-surface effects are possible also
in N-band 1D-dimensional interacting models,
e.g.\ N-leg ladders \cite{led00}. 
Interest in the field of ladders systems
has grown rapidly in the last years due to the
discovery of superconductiviy in a doped
two-leg ladder material \cite{ueh96,ric93} and due
to the fact that a rigorous weak-coupling analysis
can be carried through.

There have been a number of previous
RG investigations of 1D interacting two-band
systems. Penc and S\'olyom \cite{pen90} extended
an earlier calculation of Varma and Zawadowski \cite{var85}
and studied the Fermi-velocity 
renormalization at two-loop. Fabrizio \cite{fab93}
considered the role of the transverse hopping
$t_p$ on the stability of the Luttinger-liquid (LL)
state both for the case of particles with and 
without spin. Balents and Fisher \cite{bal96}
considered the same model we do, a general
1D model with four Fermi points. There are
several phases with $x$ gapless charge and
$y$ gapless spin modes (notation: C$x$S$y$)
in the phase diagram wich were identified
by Balents and Fisher considering the
bosonized form of the leading divergent 
running couplings. The new feature we wish
to stress is that the general phase diagram
is influenzed by the RG-flow of the Fermi vectors,
which we evaluate explicitly.
This Fermi-surface effect is especially strong
near a saddle point.

\begin{figure}[t]
\noindent
(a)\\
\centerline{
\epsfig{file=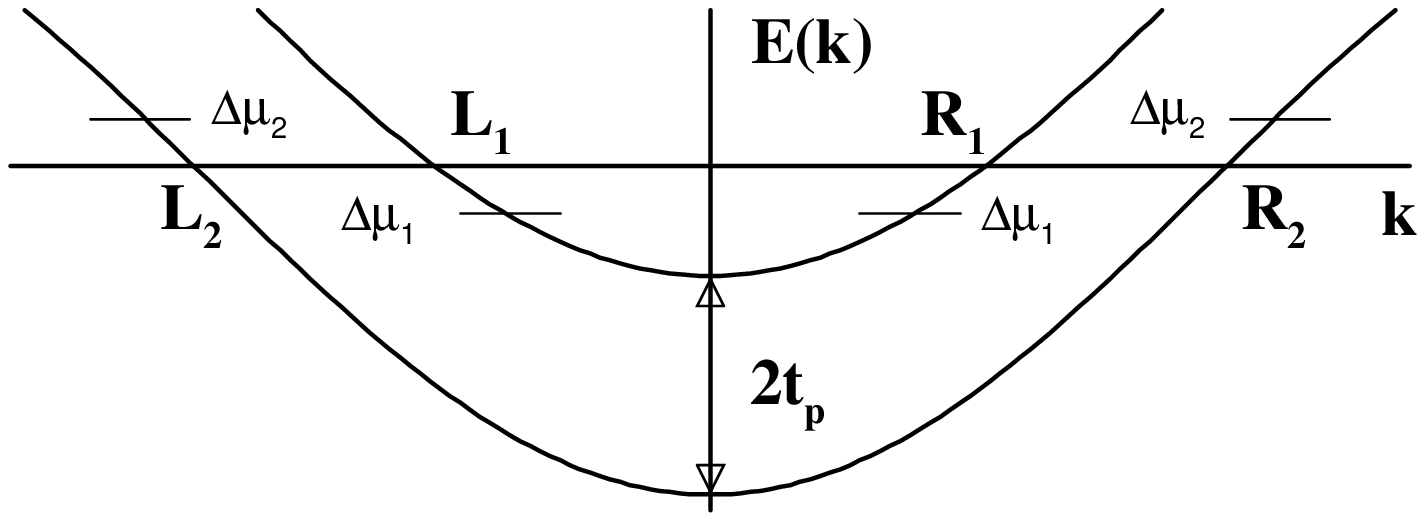,width=0.35\textwidth} 
           }
(b)\\
\centerline{
\epsfig{file=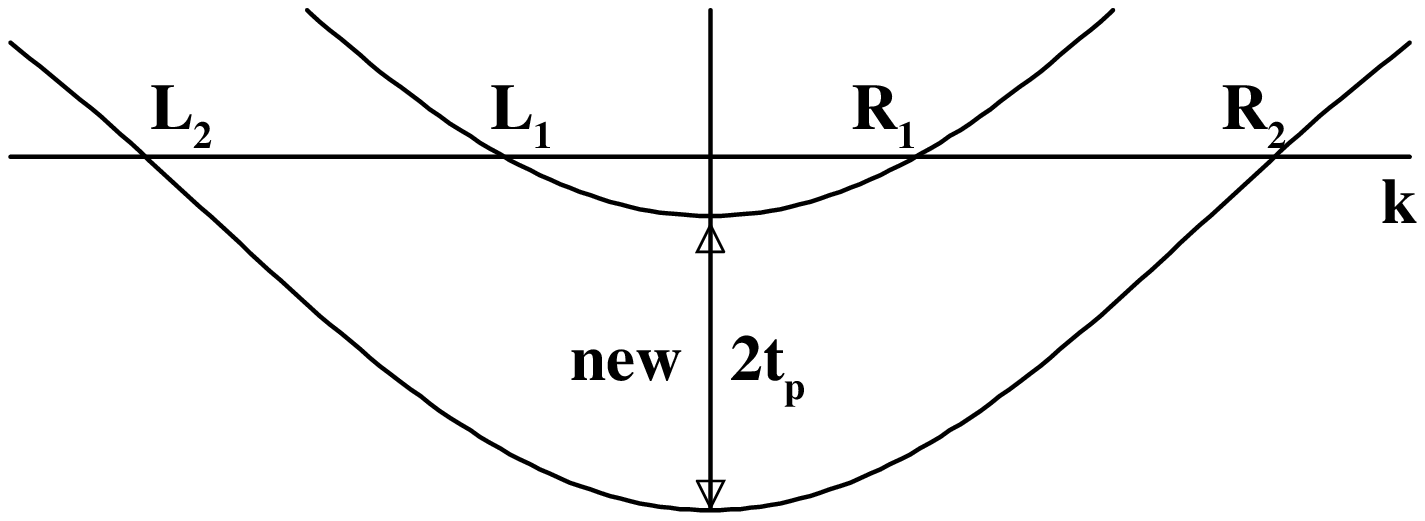,width=0.35\textwidth} 
           }
\vspace{4pt}
\caption{\label{bands2}
Dispersion relation of a Hubbard ladder with interchain hopping
$t_p$. The renormalization flow will lead, in general, to different
``chemical potentials'' for the two bands (a), which in turn
will lead to renormalized Fermi-wavevectors (b). For the case of the
Hubbard ladder, this process can be recast in a renormalization of
the interladder hopping $t_p$.
        }
\end{figure}

It is well known \cite{sha94} that the 
tadpole (or Harttree) diagram contributes to the renormalization
of the quadratic part of the original Hamiltonian,
thus leading to a renormalization of the chemical
potential in the case of a one-band model.
Alternatively, one might wish to work with a
constant density of particles. This can be 
achieved by the introduction of appropriate
counter terms to the chemical potential \cite{sha94}.
Here we would like to point out, that the tadpole
diagram leads, in general, to different
quadratic terms, which we denote by $\mu_1$ and
$\mu_2$, for the two bands in a 1D two-band model,
see Fig.\ \ref{bands2}. The renormalization of the
chemical potential is given in the 2-band case by
suitable avarage of $\mu_1$ and $\mu_2$. 
The difference of the two quadratic terms results, on the
other hand, in a renormalization of the
interchain hopping $t_p$ and thus alters the
band structure.

{\bf RG-equations} -
We will consider for a start a general
two-band model with a  linearized one-particle spectrum
around the Fermi-points $k_1$ and $k_2$
and denote with
$v_1$ and $v_2$ the Fermi-velocities
for the first band ($L_1$ and $R_1$) and
second band ($L_2$ and $R_2$) respectively.
Lateron we will specify our results to the
case of the Hubbard ladder and $t_1-t_2$
Hubbard model.
Since we will be working at one-loop, we will keep
the Fermi-velocities constant throughout the
calculation, even when the Fermi-points
$k_1$ and $k_2$ are allowed to shift.

A list of possible coupling constants is given
in Table \ref{table1}. The special couplings
$g_{2v}$, $g_w$, $g_{1v}$ and $g_s$ 
exist only for the $t_1-t_2$ Hubbard model.
The couplings listed come in two flavors, namely
for scattering between parallel and
antiparallel spins ($||$ and $\perp$)
for backward scattering,
$g_{\gamma 1j}$, and forward scattering,
$g_{\gamma 2j}$, with $\gamma=1,2,x,t$
and $j=\parallel,\perp$.
Back- and forward scattering amplitudes
for parallel spins are not independent, we choose
$g_{\gamma 1\parallel}\equiv0$.

Spin-rotational invariance implies that the
RG-flow leaves the combinations
\begin{equation}
g_{\gamma \rho} := 
-{g_{\gamma2\parallel}+g_{\gamma2\perp}\over2}~,\qquad
g_{\gamma \sigma}:=2g_{\gamma1\perp}
\label{spin_rot_inv}
\end{equation}
invariant ($\gamma=1,2,x,t$),
Here the index $\rho$ and $\sigma$
denote the charge and spin component respectively.


\begin{table}
\caption{\label{table1}Coupling constants for (a) Umklapp
and (b) normal scattering processes ($k_2>k_1$ has been assumed).
For the $t_1-t_2$ Hubbard model one has to interchange 
$R_2\leftrightarrow L_2$, compare Fig.\ (\ref{bands2})
and (\ref{phase_dia}).
         }

\begin{tabular}{c|c|c||c|c} 
label  & condition & process$^{(a)}$ & label & process$^{(b)}$ \\
\hline
\hline
$g_{2u}$ & $k_2=\pi/2$ & $L_2L_2\to R_2R_2$ &
 $g_2$ &  $L_2R_2\to L_2R_2$ \\
$g_{2v}$ & $3k_2+k_1=2\pi$ &$L_2L_2\to R_1R_2$ &
 $g_1$ &  $L_1R_1\to L_1R_1$ \\
$g_{w}$ & $3k_2-k_1=2\pi$ & $L_2L_2\to L_1R_2$ &
  $g_x$ &  $L_1R_2\to L_1R_2$ \\
$g_{xu}$ & $k_2+k_1=\pi$ & $L_2L_2\to R_1R_1$ &
  $g_t$ &  $L_2R_2\to L_1R_1$ \\
$g_{tu}$ & $k_2+k_1=\pi$ &$L_2L_1\to R_1R_2$ &
$g_s$ & $L_2R_1\to L_1L_1$ \\
$g_{1v}$ & $k_2+3k_1=2\pi$ &$L_2L_1\to R_1R_1$ &
      & ($3k_1=k_2$)\\
$g_{1u}$ & $k_1=\pi/2$ &$R_1R_1\to L_1L_1$ &&\\
\end{tabular}
\end{table}

\begin{figure}[t]
\centerline{
\epsfig{file=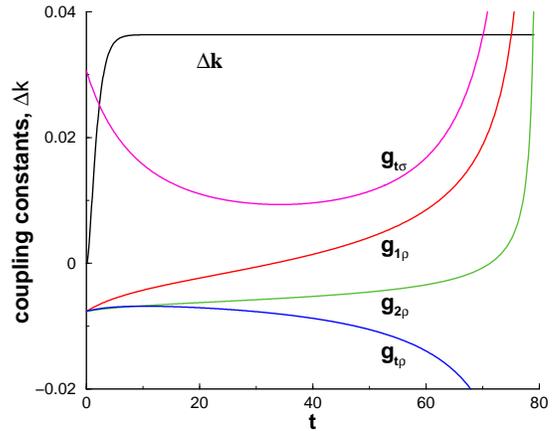,width=0.40\textwidth} 
           }
\vspace{4pt}
\caption{\label{RG_flow}
Example of the RG-flow of some selected coupling constants
for the half-filled $t_1-t_2$ Hubbard model with 
$U=0.1t_1$, $t_2=0.7t_1$, $\alpha=0.63$, $\beta=2.4$ 
(C1S0 phase, see Fig.\ \protect\ref{phase_dia}, corresponding
to $k_1=0.418\pi$, $k_2=0.918\pi$ and $v_1=1.66t_1$,
$v_2=0.43t_1$.
The initial values of the coupling constants are given by
Eq.\ (\ref{initial}). Also shown is the flow of the
Fermi-surface shift $\Delta k$ (in units of $\Lambda_0/100$).
        }
\end{figure}

The RG-equations to one-loop and momentum
cut-off $\Lambda$ are for general densities\cite{bal96}
\[
\begin{array}{rlrl}
 \dot g_{1\rho}=& \beta \left(g_{t\rho}^2+\frac{3}{16}g_{t\sigma}^2\right) &
 \dot g_{1\sigma}=&-\alpha g_{1\sigma}^2
 -\frac{\beta g_{t\sigma}^2}{2}+2\beta g_{t\rho}g_{t\sigma}\\
 \dot g_{2\rho}=& 
 \alpha \left(g_{t\rho}^2+\frac{3}{16}g_{t\sigma}^2\right) &
  \dot g_{2\sigma}=&-\beta g_{2\sigma}^2-\frac{\alpha g_{t\sigma}^2}{2}
   +2\alpha g_{t\rho}g_{t\sigma}\\
 \dot g_{x\rho}=& -\left(g_{t\rho}^2+\frac{3}{16}g_{t\sigma}^2\right) &
  \dot g_{x\sigma}=& -g_{x\sigma}^2
   -\frac{1}{2}g_{t\sigma}^2-2 g_{t\rho}g_{t\sigma}\\
 \dot g_{t\rho}=&g_{0\rho}g_{t\rho}+
 \frac{3g_{0\sigma}g_{t\sigma}}{16} &
 \dot g_{t\sigma}=&g_{0\sigma}g_{t\rho}\,+\\
 &&& 
 \left(g_{0\rho}-{g_{0\sigma}\over2}-2g_{x\sigma}\right)g_{t\sigma}~,
\end{array}
\]
where the coupling constants have been rescaled
by $1/(\pi(v_1+v_2))$, $t=-\ln(\Lambda/\Lambda_0)$
($\Lambda_0$ is the initial cut-off) and where we
have introduced
\begin{equation}
\alpha\ =\ {v_1+v_2\over2\,v_1}~,\qquad
 \beta\ =\ {v_1+v_2\over2\,v_2}
\label{ab}
\end{equation}
($1/\alpha+1/\beta=1$).
The initial values of the rescaled
coupling constants are
\begin{equation}
g_{\gamma\rho}\ =\ {-U\over 2\pi(v_1+v_2)}~,\qquad
g_{\gamma\sigma}\ =\ {2U\over \pi(v_1+v_2)}
\label{initial}
\end{equation}
for the Hubbard model.
Inclusion of Umklapp scattering at special fillings
leads to additional terms. They are 
\begin{equation}
\begin{array}{rcl}
\dot g_{1\rho}  &=& \dots\, -\,\alpha g_{1u}^2\\
\dot g_{1u}     &=& -\,4\alpha\,g_{1u}g_{1\rho}
\end{array}
\label{1u}
\end{equation}
for $g_{1u}$, where the dots in the first equation denote
the generic terms\cite{factor2}. We find 
\begin{equation}
\begin{array}{rcl}
\dot g_{t\rho}  &=& \dots\, +\,{\alpha\over2} g_s^2~,\qquad
\dot g_{t\sigma} \ =\ \dots\, -\,2\alpha\,g_s^2\\
\dot g_s     &=& g_s\,\Big(\,
-2\alpha g_{1\rho}- 3g{x\sigma}/4 + g_{x\rho} \,\Big)
\end{array}
\label{s}
\end{equation}
for $g_s$ and
\begin{equation}
\begin{array}{rcl}
\dot g_{\gamma'\rho}  &=& \dots\, -\,\alpha g_{1v}^2~,\qquad
\dot g_{1\rho} \ =\ \dots\, -\,2 g_{1v}^2 \\
\dot g_{1v}     &=& g_{1v}\,\Big(\,
-2g_{t\rho}- 2g{x\rho} - 2\alpha g_{1\rho}
    \,\Big)
\end{array}
\label{1v}
\end{equation}
for $g_{1v}$ (with $\gamma'=x,t$).
The RG-equations can be solved numerically,
we present in Fig.\ \ref{RG_flow} a typical
flow diagram for some selected coupling constants.

{\bf Phase diagram} -
Above RG-equations can be easily implemented
either for the Hubbard ladder or for the
$t_1-t_2$ Hubbard with respective
dispersion relations
\begin{equation}
\begin{array}{rcl}
E_{\pm}^{(lad)}(k)&=&\pm t_p-2t\cos(k)\\
E^{(12)}(k)&=&-2t_1\cos(k)-2t_2\cos(2k)~.
\end{array}
\label{disp}
\end{equation}
The reflection symmetry in the
Hubbard ladder (exchange of the two legs)
forbids any coupling with an odd number of
operators per band, see Tables \ref{table1} and \ref{table2}.

\begin{figure}[t]
\centerline{
\epsfig{file=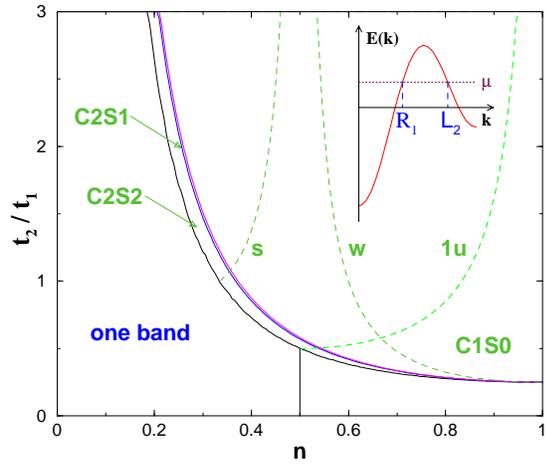,width=0.40\textwidth} 
           }
\vspace{4pt}
\caption{\label{phase_dia}
Phase diagram for the $t_1-t_2$ Hubbard model. The region denoted
`one band' is the C1S1 LL-phase with one band empty, $n=0.5$
denotes half-filling.
Note that the `1u' phase-line is continuous\protect\cite{factor2}.
Inset: An illustration of the dispersion relation.
        }
\end{figure}

In Fig.\ \ref{phase_dia} we present the phase-diagram for
the  $t_1-t_2$ Hubbard model, for which
the control-parameter $\beta$ takes the form
\begin{equation}
\beta\ =\ {v_1+v_2\over2v_2}\ =\ 
{\sin(k_1)+\sin(k_2)\over2\sin(k_2)}~,
\label{beta}
\end{equation}
At half
filling $k_2-k_1=\pi/2$.
Depending on $\beta$, various coupling constants
might scale to strong coupling, the divergence is
in general of the form $\sim g^*/(t^*-t)$. In
Table\ \ref{table2} a complete list of the
diverging coupling constants is given. The
differences between the Hubbard ladder and
the $t_1-t_2$ Hubbard model, which stem from
the different geometries of the two Fermi-seas,
are pointed out in Table\ \ref{table2}.

The Umklapp scattering $g_{2u}$ at $k_2=\pi/2$ is
not active in the $t_1-t_2$ Hubbard model, 
since $k_2>\pi/2$  for this model.
$g_{2v}$ is allowed in the $t_1-t_2$ Hubbard model
and would leads to a C1S0 phase if (a)
$3k_2+k_1=2\pi$ and (b) $\beta>2.4$. It turns out,
that it is not possible to satisfy
both conditions at the same time.
Inspecting the fix-point Hamiltonian
and bosonizing the diverging coupling 
constants \cite{bal96}, see Table \ref{table2}, 
one finds that the number of gapless
charge and spin modes are
C1S2 (1u,2u), C1S0 (1v,2v), C1S1 (s,w) and
C0S0 (xu,tu).

{\bf RG of Fermi-wavevectors} -
The renormalization equations for the
local chemical potentials are given by
the tadpole diagram:
\begin{equation}
\dot\mu_i\ =\ \Lambda(v_1+v_2)\left(g_{i\rho}+g_{x\rho}\right)~,
\qquad (i=1,2)~.
\label{dot_mu_i}
\end{equation}
Note that the cut-off $\Lambda$ occurs on the 
right-hand side of Eq.\ (\ref{dot_mu_i}) and that
the $\dot\mu_i$ are linear in the coupling constants.
Let's consider now a renormalization step, we write
\begin{equation}
\Delta\mu_i\ =\ \Delta\mu_i^{0}\,+\, \Delta\mu_{ct}~,
\label{Delta_mu}
\end{equation}
where the counter-term $\Delta\mu_{ct}$ is introduced
in order to enforce particle number conservation.

\begin{table}
\caption{\label{table2}Diverging coupling constants
in all possible phases, $\kappa=\mbox{max}(\alpha,\beta)$.}
\begin{tabular}{cccc}
phase & conditions & \multicolumn{2}{c}{divergent couplings} \\
      &            & $\to(-\infty)$ & $\to(+\infty)$ \\
\hline
C2S2 & $\kappa>4.8$     & (none)   &   (none)\\
C2S1 & $4.8>\beta>4.3$  &  $g_{1\sigma}$ & (none)\\
C2S1 & $4.8>\alpha>4.3$ &  $g_{2\sigma}$ & (none)\\
C1S0 & $4.3>\kappa$    & $g_{t\rho},g_{x\rho},g_{1,2\sigma}$ &   
 $g_{t\sigma},g_{1,2\rho},(g_{x\sigma})^\ddag$ \\
\hline
 1u  &                  & $g_{1\rho}$   &   $g_{1u}$ \\
 2u$^\dagger$  &        & $g_{2\rho}$   &   $g_{2u}$ \\
\hline
1v$^*$,2v$^{*\dagger}$ &  $\kappa>2.4$ & $g_{t\rho},g_{x\rho},g_{1,2\sigma}$ &
                                $g_{t\sigma},g_{1,2\rho},g_{x\sigma}$  \\
\hline
s$^*$  &             & $g_{x\rho},g_{1\rho},g_{t\rho}$ &
                       $g_{s},g_{2\rho}$   \\
w$^*$  &             & $g_{x\rho},g_{2\rho},g_{t\rho}$ &
                       $g_{w},g_{1\rho}$   \\
\hline
xu$^\dagger$,tu$^\dagger$& & $g_{tu\rho}$ &
                             $g_{xu},g_{tu\sigma}$  \\
 &   & $g_{t\rho},g_{x\rho},g_{1,2\sigma}$ &
                                $g_{t\sigma},g_{1,2\rho},g_{x\sigma}$  \\
\end{tabular}
$^*$Does not occur in the ladder due to transversal momentum conservation.\\
$^\dagger$ Does not occur in the $t_1-t_2$ model due to
conditions on $k_1$ and $k_2$.\\
$^\ddag$ Subleading divergence $\sim g_{x\sigma}^*/\sqrt{t^*-t}$. 
\end{table}

The counter term
is compensated by subtracting the corresponding
term from the interaction part of the Hamiltonian.
The size of $\Delta\mu_{ct}$ is determined for the
Hubbard ladder by the condition 
($\Delta k_i v_i= \Delta\mu_i^0$)
\begin{equation}
\Delta k_1\ =\ -\Delta k_2~,\qquad
v_2\Delta\mu_1^0\ =\ -v_1\Delta\mu _2^0
\label{cond_Delta_k}
\end{equation}
(for the $t_1-t_2$ Hubbard model
$\Delta k_1=\Delta k_2$ etc.), which leads to
\begin{equation}
\begin{array}{rcl}
\displaystyle
\Delta\mu_i^0 &=& (-1)^i\,{v_i\over v_1+v_2}
     \big( \Delta\mu_2-\Delta\mu_2\big)\\ \displaystyle
\Delta\mu_{ct} &=& {v_1\over v_1+v_2} \Delta\mu_2
                  +{v_2\over v_1+v_2} \Delta\mu_1~.
\end{array}
\label{mu_terms}
\end{equation}
Eq.\ (\ref{mu_terms}) and
Eq.\ (\ref{dot_mu_i}) lead then to
\begin{equation}
\dot k_i\ =\ (-1)^i \Lambda
\left(g_{2\rho}-g_{1\rho}\right)~,
\label{dot_k_i}
\end{equation}
which describes the flow of the Fermi-wavevector.
Since the initial values for
$g_{1\rho}$ and $g_{2\rho}$ are identical,
see Eq.\ (\ref{initial}), a non-trivial
renormalization of the Fermi-wavevector occurs
only when $v_1\ne v_2$, i.e.\ when
$\alpha\ne\beta$. For the Luttinger-liquid phase
C2S2 the limit
\begin{equation}
\Delta k_i = \lim_{t\to\infty}\int_0^t dt'\,
\dot k_i(t')  \label{Delta_k_i}
\end{equation}
is well defined. In the other phases there
is normally a plateau in
$\Delta k_i(t) = \int_0^t dt'\, \dot k_i(t')$,
due to the decreasing cut-off $\Lambda$ in
the expression (\ref{dot_k_i}) for
$\dot k_i$ which can be considered to define
an approximate $\Delta k_i$, see Fig.\ \ref{RG_flow}.
 
Let us consider the flow of the Fermi-wavevector
in some more detail for the Hubbard ladder, where
$k_1<k_2$ and $v_1<v_2$. The relation $\alpha>\beta$
then leads generally to $g_{2\rho}>g_{1\rho}$
and consequently to
\begin{equation}
\Delta k_1 \ < \ 0~,\qquad 
\Delta k_2 \ > \ 0~.
\label{sign_k_12}
\end{equation}
The smaller Fermi-sea tends to empty.

In the limit $U\to0$ the renormalization
of the coupling constants tends to zero in
the C2S2 phase, and one can obtain an asymptotic
rigorous expression for the total shift 
$\Delta k=-\Delta k_1$
using ($i=1,2$)
\[
g_{i\rho}(t)\ \approx\ g_{i\rho}(0)
+ \dot g_{i\rho}(0)\,t~,\qquad
\dot g_{i\rho}(0)\ =\ 
{U^2/2\over v_1+v_2} {1\over\pi^2v_{3-i}} 
\]
and Eq.\ (\ref{initial}) for $g_{i\rho}(0)$.
With (\ref{dot_k_i}) and (\ref{Delta_k_i})
we obtain
\begin{eqnarray}
\nonumber\Delta k &\approx& 
{\Lambda_0U^2/2\over\pi^2(v_1+v_2)}
\left({1\over v_1}-{1\over v_2}\right)\\
&=& 
{U^2\over\pi^2}\,
{\alpha-\beta\over(v_1+v_2)^2}\,\Lambda_0~.
\label{Delta_k}
\end{eqnarray}
The Fermi-surface renormalization effects become especially
strong near a saddle point, where $v_1\ll v_2$, 
i.e.\ $\alpha\gg\beta$. Eq.\ (\ref{Delta_k}) has been derived
for the case of the Hubbard ladder. For the $t_1-t_2$
Hubbard model one has just to interchange 
$\alpha\leftrightarrow\beta$, both Fermi-points shift then
in the same direction (to larger values), see the
inset of Fig.\ \ref{phase_dia}.
%

{\bf Discussion} -
In above discussion of the Fermi-surface renormalization we
have kept the Fermi-velocities $v_1$ and $v_2$. At two-loop,
the Fermi-velocities are renormalized, an effect $O(U^3)$.
The change of $\delta k$ is, on the other hand,\
quadratic in $U$ and leads therefore (for the ladder)
to an additional reduction in $v_1$, which is
$O(U^2)$. This reduction in $v_1/v_2$ would lead
to a further enhancement of
$\Delta k$ in a self-consistent treatment. We do not, however,
expand this point here further, since it would involve
a somewhat uncontrolled mixture of high- and low-energy
scales.

It is of interest, however, to study the stability of the
saddlepoint itself. This question has been studied
extensively for the case of $N$-patch models of the 
2D Hubbard model. Honerkamp {\it et al.} \cite{fur98}
have found, that particles tend to leave the saddle-point
region (also called `hot spots'), in accordance to our
result Eq.\ (\ref{Delta_k}). Gonzalez {\it et al.}
\cite{gon00,gon97}
have found that the saddle point attracts the Fermi-level
both from above and from below. 

We have therefore considered the
case when the Fermi-level is exactly at the bottom
of the first band in the Hubbard ladder. In this case the
usual RG-approach has to be modified.
We have followed the approach proposed by Balents
and Fisher \cite{bal96} (an expansion in the
curvature) and find with $v_1=0$
\begin{equation}
\dot\mu_1\ =\ \Lambda v_2 g_{x\rho}\qquad
\dot\mu_2 = \Lambda v_2 g_{2\rho}~.
\label{mu_vH}
\end{equation}
As the initial values of $g_{x\rho}$ and
$g_{2\rho}$ are identical for the Hubbard model,
see Eq.\ (\ref{initial}), one has to consider the
flow of $g_{x\rho}$ and $g_{2\rho}$, which is,
due to the lack of scale-invariance, explicitly
cut-off dependent. One finds \cite{note_vH}
$g_{2\rho} > g_{x\rho}$, i.e.\
the Fermi-level is pushed below the
bottom of the first band. The saddle point is
therefore pushed above the Fermi-level by the RG-flow.


{\bf Mott-Hubbard transition} -
A Mott-Hubbard transition occurs in the half-filled
one-dimensional $t_1-t_2$ model as a function of 
interaction strength $U$. For $t_2>t_1/2$, that is 
for the case of four Fermi-points, this transition
occurs at finite values of $U$. Aebischer {\it et al.}
estimated \cite{aeb01}
a critical $U_c\sim2.67t_1$ for $t_2=0.7t_1$. The question
we would like to ask now is: How large is the Fermi-surface
shift $\Delta k$ for $U=U_c$? We cannot answer this question
within our weak-coupling approach, but we can estimate the
order of magnitude of $\Delta k$ by assuming formula
Eq.\ (\ref{Delta_k}) to hold up to the Mott-Hubbard transition.
We obtain
\begin{equation}
\Delta k\big|_{U=2.67 t_1}^{t_2=0.7t_1}\ \approx\ 0.4\,\Lambda_0
\ =\ 0.4\,(\pi-k_2)~,
\label{shift_MH}
\end{equation}
where we have taken $\pi-k_2$ as a physical relevant 
initial cut-off.
This estimate indicates that the Fermi-wavevector
is shifted by a substantial fraction towards the Brillouin
zone edge near the Mott-Hubbard transition. We note that
the real $\Delta k$ might be even larger since we have
neglected in this estimate the renormalization of the
Fermi-velocity which, as we have discussed further above,
would enhance the effect. In principle it would be possible
that the Mott-Hubbard transition coincides with 
$k_2\to\pi$ for $U\to U_c$. Physically this
would correspond to an activation of the one-band
Umklapp scattering $g_{1u}$ at the transition point.


{\bf Conclusions} -
We have derived and discussed the RG-equations for the
Fermi-wavevectors in one-dimensional models with four
Fermi-points. We find that the renormalization of the
Fermi-wavevectors is quadratic in the interaction, suggesting 
substantial effects at strong coupling 
and near the Mott-Hubbard transition.  We have pointed out, that
the renormalization of the Fermi-wavevector is of especial
importance near a saddle point, and that it leads to 
changes, $\Delta v_i$, in the Fermi-velocities 
quadratic in the coupling constant $U$, 
whereas the usual two-loop contributions to
$\Delta v_i$ are cubic in $U$.

In addition we have discussed the phase diagram of the
1D $t_1-t_2$ Hubbard model and found a new phase
(C1S1,s/w).

{\bf Acknowledgments} -
This work was partially supported by the DFG. One of us 
(C.G.) would like to thank W. Metzner for valuable 
discussions.




\begin{thebibliography}{99}
\bibitem{factor2} Note the factor two difference of
                  Eq.\ (\protect\ref{1u}) with respect to
		  the corresponding Eq. (3.1) in
		  Ref.\ \protect\cite{bal96}, which we
		  believe to be erroneous. As a consequence
		  of this difference the line of existence
		  for the ``1u'' phase (and for the corresponding
		  ``2u'' phase) are not broken in 
		  Fig.\ \protect\ref{phase_dia}.
\end{thebibliography}
\end{document}